\documentclass[3p,times,procedia,preprintnumbers,amsmath,amssymb,nofootinbib,epsfig,bmamsfonts,yfonts,superscriptaddress]{elsarticle}
\usepackage{color}

\usepackage{amssymb}
\usepackage{amsmath}
\usepackage{ulem}

\makeatletter
\newcommand{\xRightarrow}[2][]{\ext@arrow 0359\Rightarrowfill@{#1}{#2}}
\makeatother

\usepackage[figuresright]{rotating}

\begin{document}

\begin{frontmatter}

\title{Opacity dependence of elliptic flow in kinetic theory}

\author{Aleksi Kurkela$^{1,2}$, Urs Achim Wiedemann$^{1}$ and Bin Wu$^{1}$}

\address{$^1$ Theoretical Physics Department, CERN, CH-1211 Gen\`eve 23, Switzerland\\
$^2$ Faculty of Science and Technology, University of Stavanger, 4036 Stavanger, Norway}

\begin{abstract}
The observation of large azimuthal anisotropies $v_n$ in the particle spectra of proton-proton (pp) and proton-nucleus (pA) collisions challenges fluid dynamic interpretations of $v_n$, 
as it remains unclear how small collision systems can hydrodynamize and to what extent hydrodynamization is needed to build up 
$v_n$. Here, we study in a simple kinetic theory how the same physics that leads to hydrodynamization in large systems represents itself in small systems. 
We observe that one third to one half of the elliptic flow signal seen in fully hydrodynamized systems can be built up in collisions that extend over only one mean free path $l_{\rm mfp}$ and 
that do not hydrodynamize. This is qualitatively in line with observing a sizeable $v_2$ in $pp$ collisions  for which other characteristics of soft multi-particle production 
seem well-described in a free-streaming picture.
We further expose a significant system size dependence in the accuracy of hybrid approaches that match kinetic theory to viscous fluid dynamics.   
The implications of these findings for a reliable extraction of shear viscosity are discussed.

\end{abstract}



\end{frontmatter}
\vspace{0.5cm}

{\bf Introduction}.
Ultra-relativistic nucleus-nucleus (AA), proton-nucleus (pA) and proton-proton (pp) collisions display remarkably large signatures of collectivity, in particular in the hadronic transverse momentum spectra and their azimuthal asymmetries $v_n$~\cite{ALICE:2011ab,Abelev:2014mda,Khachatryan:2015waa,Aaboud:2017acw,Adare:2014keg,Adamczyk:2015xjc}. 
To infer the properties of the ultra-dense and strongly expanding QCD matter in the collision region from these data, a dynamical modelling of collectivity is indispensable. From comparing fluid dynamic models to data of large (AA) collision systems, one generally determines matter properties consistent with a perfect fluid that exhibits minimal dissipation (having minimal shear viscosity over entropy ratio, $\eta/s$)~\cite{Heinz:2013th,Romatschke:2017ejr}. In marked contrast, the standard implementation of soft multi-particle production in general-purpose event generators~\cite{Buckley:2011ms} of pp collisions implements a
free-streaming picture according to which outgoing quanta do not interact with each other. Kinetic transport theory is of particular interest in this context since it can in principle interpolate between 
the limiting cases of free-streaming and fluid-dynamic behavior. Indeed, transport models have been demonstrated to account for the signals of collectivity in pA and AA collisions with material properties 
that allow for a significant mean free path, thus exhibiting non-minimal 
dissipation~\cite{Borghini:2010hy,Xu:2011fi,Xu:2011jm,Uphoff:2014cba,He:2015hfa,Koop:2015wea,Greif:2017bnr,Kurkela:2018ygx,Borghini:2018xum}. 

Despite these recent advances in applying transport theory to hadronic collisions, our dynamical understanding of the system size dependence of collectivity remains incomplete. 
Even elementary questions---such as: what is the minimal
(maximal) size over which a hadronic collision system needs to extend to exhibit fluid-like properties (free streaming)?---still await systematic exploration. To address these questions, 
the need for developing even more realistic and more complex simulation tools of pp, pA and AA collisions is widely acknowledged. But to better understand generic, model-unspecific characteristics
of how collectivity is built up in collision systems of sizeable mean free path, one should {\it also} explore the opposite direction:  One should also study in isolation
particularly simple formulations of kinetic theory that are not embedded in the multi-layered reality of fully realistic simulation packages. Such formulations ideally depend on as few model parameters as possible, 
and they are ideally free of model-specific assumptions about, {\it e.g.}, hadronization or about detailed dynamical approximations entering the collision kernel. 

Many microscopic models with boost-invariant longitudinal dynamics satisfy hydrodynamic constitutive equations in situations significantly out of equilibrium, an observation dubbed ``hydrodynamization without thermalization''\cite{Heller:2011ju,Chesler:2010bi,Wu:2011yd,Heller:2013oxa, Kurkela:2015qoa,Keegan:2015avk,Heller:2016rtz}. Ultra-relativistic pp, pA and AA collisions realize such out-of-equilibrium scenarios 
since they are initiated with a highly anisotropic momentum distribution. However, whether the process of hydrodynamization is completed or not may depend significantly 
on the transverse extent and lifetime of the collision system.  Here, we analyze systematically over the entire range of physically relevant system sizes how a particularly simple, one-parameter kinetic
theory can account for the 
requirements of realizing a close-to-hydrodynamic behavior on time scales comparable with a nuclear radius while supporting a close-to free-streaming picture in minimum bias pp collisions, and exhibiting for small but increasing system sizes a rapid onset of sizeable signals of collectivity.  The simple kinetic theory employed here is based only on arguably generic assumptions 
about the isotropizing character of rescattering phenomena, and it will be shown to exhibit with increasing system size an increasing degree of hydrodynamization. The model thus provides a simple testbed for understanding to what extent signals of collectivity can arise in small systems with negligible or partial hydrodynamization, and how they can grow with increasing system size. 
In addition, this approach allows one to quantify the system-size dependent uncertainties that arise from interfacing a pre-hydrodynamic evolution based on kinetic transport with a subsequent fluid dynamic description. 

In general, any kinetic transport formulation assumes a scale separation between the typical size of the wave-packet of particle-like excitations and their mean free path. This assumption 
is not realized, {\it e.g.}, in models of strongly coupled liquids formulated in the limit of strong coupling with the gauge/gravity conjecture. This assumption would be supported, however, 
by any evidence for the dominance of free-streaming in a small collision system, since free-streaming over some finite extent translates trivially to a lower bound on the mean free path. 
Moreover, any finite mean free path implies non-minimal dissipative material properties and thus translates to a non-minimal constraint on  $\eta/s$. In this sense, establishing a unified 
dynamical description of collectivity valid from AA via pA to the smallest pp collision systems has the potential of providing  a complementary constraint on $\eta/s$  from the system size 
dependence of $v_n$ data.

{\bf Kinetic transport: the model}.
Our study focusses on azimuthal asymmetries $v_n$ of the transverse energy $d E_\perp$  that are trivially obtained from 
those of measured particle spectra $dN$, 
\begin{equation}
 \frac{d E_\perp}{d \eta_s d \phi} \!\equiv\!\! \int\!\! dp^2_\perp   \frac{p_\perp\, d N}{dp^2_\perp d \eta_s d \phi} \!
=\! \frac{d E_\perp}{2\pi d \eta_s}\! \left(\!1\! +\! 2\sum_{n=1}^\infty v_n\cos(n\, \phi) \right) \, .
\label{eq1}
\end{equation}
In comparison to $dN$, an analysis of $dE_\perp$ is not complicated by the potentially confounding effects of hadronization. We calculate
$d E_\perp$ by evolving the energy-momentum tensor $T^{\mu\nu}$ of the system to late times. To this end, we write
$T^{\mu \nu} = \int_{-1}^{1}\frac{d v_z}{2} \int \frac{d\phi}{2\pi} v^{\mu} v^{\nu} F$ in terms of the 
first momentum moment $F(\vec x_\perp,\Omega,\tau) = \int\frac{4 \pi p^2 d p}{(2\pi)^3} p f$ of the distribution function $f$.
Here, $p$ is the modulus of the three-momentum, and we use normalized momenta $v_\mu \equiv p_\mu/p$ with $p_\mu\, p^\mu = 0$ and $v^0 = 1$. 
The two-dimensional angular orientation $\Omega$ of the momentum can be written in terms of the azimuthal angle $\phi$ and the normalized longitudinal 
momentum component $v_z$. For massless boost-invariant kinetic transport in the slice of central spatial rapidity $\eta_s = 0$, the evolution
equation for $F$ reads~\cite{Kurkela:2018ygx}
\begin{equation}
\partial_\tau F + \vec{v}_\perp \cdot \partial_{\vec{x}_\perp} F - \frac{v_z}{\tau}(1-v_z^2) \partial_{v_z} F + \frac{4 v_z^2}{\tau}F = -C[F] \, .\label{eq2}
\end{equation}
For the collision kernel $C[F]$, we use 
the \emph{isotropization-time approximation} (ITA)
\begin{equation}
-C[F] = -\gamma \varepsilon^{1/4}(x) [-v_\mu u^{\mu} ] (F - F_{\rm iso})\, ,
\label{eq3}
\end{equation}
where  $\varepsilon$ is the local energy density and 
$F_{iso}(\vec x_\perp,\Omega,\tau)   = \frac{\varepsilon(\vec x_\perp,\tau)}{(-u_\mu v^{\mu})^4}$ is the
isotropic distribution in the local rest frame $u^\mu$ given by the Landau matching condition, $u^\mu T_{\mu}^{\, \, \,\nu}  = - \varepsilon u^\nu$. The ITA  is closely related to the relaxation time approximation. 
We emphasize, however, that for observables constructed from $T^{\mu\nu}$, it is not necessary to specify the 
momentum-dependence of $C[F]$. Eq.~(\ref{eq3}) is solely based on the mild assumption that any 
system evolves towards an isotropic distribution and that this can be characterized for $p$-integrated quantities
by a single isotropization time $l_{\rm mfp} \sim \left(\gamma \varepsilon^{1/4}\right)^{-1}$, set by the only model
parameter $\gamma$. 
The ITA has been studied extensively in the hydrodynamical limit and its transport coefficients
are known: $\tau_\pi = (\gamma \varepsilon^{1/4})^{-1}$ and kinetic shear viscosity 
$\frac{\eta}{s T} =  \frac{1 }{\gamma \varepsilon^{1/4}}\frac{1}{ 5}$. While our calculations depend only on the combination $\frac{\eta}{s T}$ and do not depend on $\frac{\eta}{s}$ independently, the
latter can be determined once the equation of state relating energy density and temperature is specified. (If one chooses $\varepsilon \approx 13 T^4$, as motivated by lattice results, one finds $\eta/s \approx 0.11/\gamma$.) We note that the thermal equilibrium distribution that enters the relaxation time approximation (RTA) is a special choice of an isotropic distribution. Therefore, the RTA and ITA dynamics for 
$T^{\mu\nu}$ are identical while the ITA does not assume relaxation to a local thermal equilibrium. 
The ITA is found to reproduce the $T^{\mu\nu}$ evolution of the QCD weak coupling effective kinetic theory~\cite{Arnold:2002zm} within $\sim$ 15\% \cite{Heller:2016rtz}.
However, the following does not assume that the collision kernel is dominated by perturbative physics.

\begin{figure*}[t]
 \includegraphics[width=0.3\textwidth]{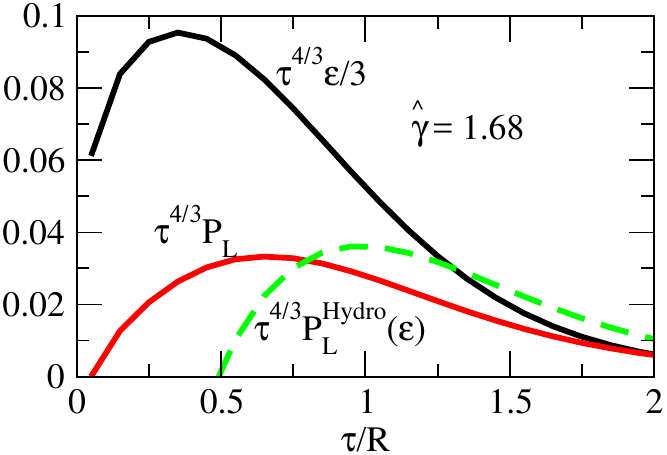}
\includegraphics[width=0.3\textwidth]{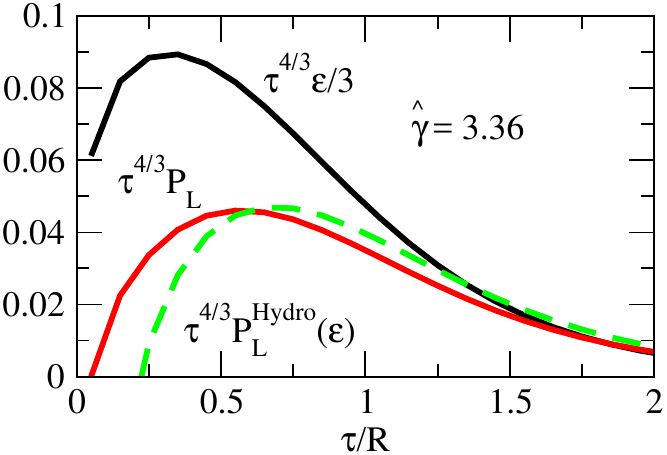}
\includegraphics[width=0.3\textwidth]{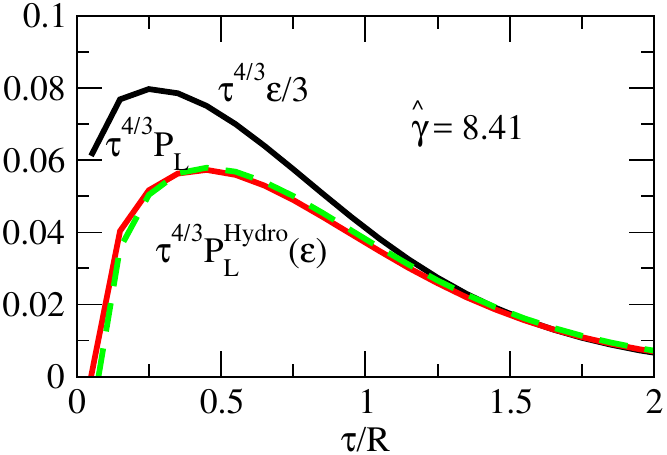}
\caption{Time evolution of the energy density $\tau^{4/3}\varepsilon$ (black) and 
the longitudinal pressure $\tau^{4/3}P_L$ (red), measured at $r=0$ and compared to  the 1st order hydrodynamic constitutive equation
(green) for different values of transverse system size $\hat\gamma$.
}
\label{fig2}
\end{figure*}

Azimuthal asymmetries $v_n$ in the final momentum distributions arise from azimuthal
eccentricities $\epsilon_n$ in the initial spatial distribution. To choose a longitudinally boost-invariant initial condition
that shares pertinent phenomenologically relevant features, we assume at each point in space an azimuthally isotropic momentum distribution with 
maximal anisotropy in the longitudinal component
($\propto \delta(v_z)$). For spatial distributions, we choose an azimuthally isotropic Gaussian density profile distorted by eccentricities 
$\epsilon_n$. Focussing for simplicity on the second harmonic, we write
\begin{equation}
F(\vec x_\perp, {\Omega}, \tau_0) = 2\varepsilon_0\, \delta(v_z)\, 
\exp\left[ -\frac{r^2}{R^2}  \right] \left(1 - {\epsilon_2} \frac{r^2}{R^2} \cos 2\theta \right)\, ,
\label{eq4}
\end{equation}
{with spatial azimuthal angle $\theta$ and radial coordinate $r$}. 
The normalization of (\ref{eq4}) corresponds to an initial central energy density $\varepsilon(\tau_0,r=0) =
\varepsilon_0$. We take $\tau_0\rightarrow 0$ keeping $\varepsilon_0\tau_0$ fixed. Then, evolving this initial condition (\ref{eq4}) with the kinetic theory (\ref{eq2}), dimensionless observables
can depend only on opacity $\hat \gamma =R^{3/4}\gamma (\, \varepsilon_0 \tau_0)^{1/4}$. This opacity may be thought of as measuring the 
transverse system size $R$ in units of mean free path at the time $\tau=R$ at which collectivity is built up, 
$\hat \gamma = R/l_{\rm mfp}(\tau=R)  \approx R \gamma \left(e\, \varepsilon(\tau=R,r=0 \right) )^{1/4} ${, where the latter equivalence is exact for a free streaming system%
\footnote{
{Here the Euler's constant $e$ arises from the time evolution of the central density in a free-streaming system $\tau \varepsilon(\vec x_\perp =0, \tau) = \tau_0 \varepsilon_0 e^{-t\tau^2/R^2}+\mathcal{O}(\gamma, \epsilon_2)$}}}.
From previous studies of this kinetic theory to first order in $\hat\gamma$, {\it i.e.}, for small system sizes,
we know already that all linear and non-linear structures observed in the azimuthal anisotropies 
$v_n$ arise, and that  $v_2/\epsilon_2 = 0.212 \hat{\gamma}$~\cite{Kurkela:2018ygx}.

Note that in physical collision systems, the opacity $\hat{\gamma}$ can be varied either by changing the geometrical size of the system $R$ or by changing the mean free path by varying the density $\varepsilon_0 \tau_0$. In central and semi-central heavy ion collisions the geometrical system size can be controlled by selection of centrality classes. In contrast, in pp collisions, one expects that the change in geometrical size plays a lesser role but $\hat{\gamma}$ may still be varied by multiplicity selection leading to denser or more dilute systems.

{\bf Kinetic transport: non-perturbative solution and results}.
The first result reported here is that we have developed a novel approach for solving the kinetic theory (\ref{eq2}) exactly to all orders in $\hat\gamma$ and thus for collision systems of any opacity. We do so by discretizing the transport equation (\ref{eq2}) in comoving coordinates that leave the distribution $\tilde F$ of free streaming particles unchanged as a function of time, and evolve it in time
numerically
\begin{equation}
\partial_\tau \tilde F_n(\tilde x_\perp,\phi,\tilde v_z,\tau) = -\frac{e^{-i n (\phi-\theta)}}{\Delta^4}C_n[\Delta^4 e^{ i n (\phi-\theta)} \tilde F_n],
\end{equation}
with
\begin{equation}
\vec x_\perp = \tilde x_\perp - \frac{\hat v_\perp}{\sqrt{1-\tilde v^2_z}}(\tau_0 - \tau\Delta), 
\quad
v_z= \frac{\tau_0}{\tau\Delta}\tilde v_z.
\end{equation}
Here, $\Delta=\sqrt{1-\tilde v_z^2 + (\tau_0/\tau)^2 \tilde v_z^2}$ and $F_n=e^{i n (\phi-\theta)}\Delta^4 \tilde F_n$ and $C_n$ correspond to the $n$th harmonic of the distribution function and the appropriately linearized collision kernel for $F_n$. A detailed description of the numerical method of solving (\ref{eq2}) will be given in Ref.~\cite{Kurkela:2019kip}.

To delineate the physically interesting parameter range for our study, we first determine the range of opacities which correspond to negligible, partial or almost complete hydrodynamization. To this end, 
we compare at the center $r=0$ of the collision, where transverse velocity is absent, the results of transport theory to the first order 
viscous constitutive equation $P^{hydro}_L = \frac{\varepsilon}{3} \left[1-\frac{16}{3} \frac{\eta}{s\, T} \left( \frac{1}{\tau}-\partial_r u_r \right) \right]$.
With increasing system size and evolution time, fluid dynamic expectations are seen to coincide better and better with transport results, see Fig.~\ref{fig2}. The kinetic theory (\ref{eq2})-(\ref{eq4}) shows hydrodynamization---that is, approximate overlap of the green and red curves in Fig.~(\ref{fig2})---for $\tau \gtrsim R/\hat \gamma$. Consistent with many recent studies \cite{Keegan:2015avk,Kurkela:2015qoa,Heller:2016rtz}, this takes place prior to thermalization, $P_L \sim \varepsilon/3$. 

For the following discussion of $v_2/\epsilon_2$, it is useful to rephrase the finding of Fig.~\ref{fig2} in terms of the properties that the collision system possesses during the typical time $\tau \sim R$ 
over which the signal for $v_2/\epsilon_2$ is predominantly built up in kinetic theory. 
Fig.~\ref{fig2} indicates then that during the timescale over which the flow is built up, the system may be characterized as not hydrodynamized for 
$\hat \gamma \leq 1.5$, as partially hydrodynamized for $1.5 \leq \hat \gamma \leq 4$, and as almost completely hydrodynamized for  $\hat \gamma \geq 4$. It should be understood that this is only a rough
characterization based on the differences between green and red curves in Fig.~\ref{fig2}, but it will be of help for discussing in the following the results of transport theory in 
qualitatively different dynamical regimes.  

\begin{figure}
\includegraphics[width=0.45\textwidth]{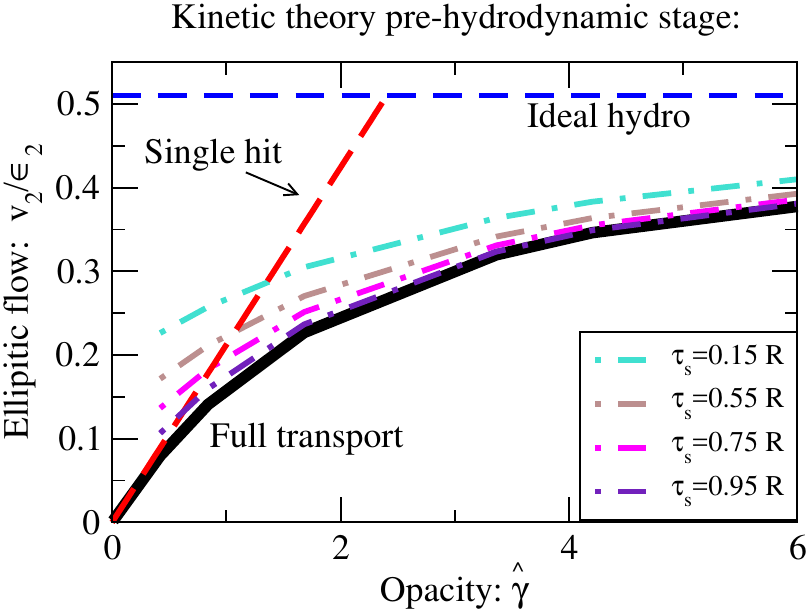}
\hfill
\includegraphics[width=0.45\textwidth]{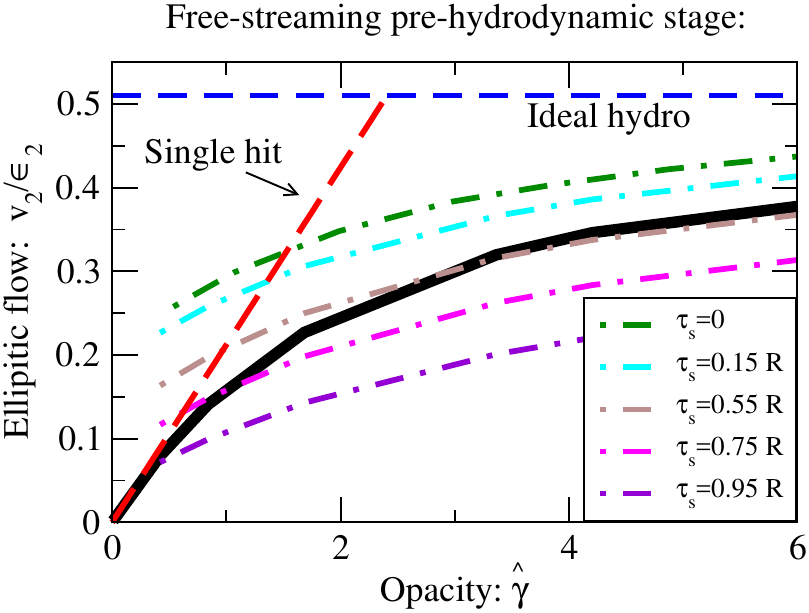}
\caption{The linear response coefficient $v_2/\epsilon_2$ as a function of $\hat\gamma =R^{3/4}\gamma (\, \varepsilon_0 \tau_0)^{1/4} = R/l_{\rm mfp}$. The thick black line is the full (all orders in $\hat\gamma$)
result obtained from evolving the kinetic theory (\ref{eq2}), (\ref{eq3}) up to arbitrarily late times. The red dashed line (single hit) is the corresponding result to first order in $\hat\gamma$. In the range of $\hat\gamma$ in which these two lines are approximately equal, the response coefficient is build up by up to $O(1)$ scatterings per particle. The dash-dotted lines correspond to multistage simulations where viscous fluid dynamics 
(with $\eta/s$ set consistently with $\gamma$) is interfaced with (a) kinetic transport (left hand side) and (b)  free-streaming (right hand side) at switching time $\tau_s$. The theoretical upper limit of $v_2/\epsilon_2$
is obtained from evolving the initial conditions with non-viscous ideal fluid dynamic (blue dashed line). }
\label{fig1}
\end{figure}

One of the main novel results of this work is the thick black curve in Fig.~\ref{fig1}. It shows how the signal strength $v_2/\epsilon_2$ calculated from the kinetic theory (\ref{eq2}), (\ref{eq3}) 
builds up smoothly over the entire physically relevant range of system sizes including systems with negligible, partial or almost complete hydrodynamization.
The full solution for $v_2/\epsilon_2$ approaches the analytically known~\cite{Kurkela:2018ygx}  first order  expression for small 
$\hat\gamma$ (see the `single-hit'  curve in Fig.~\ref{fig1}). From a technical point of view, this is a useful consistency check for the numerical accuracy of our solution. The related physics message
is that in the range in which the single hit line agrees approximately with the full transport result, the signal strength $v_2/\epsilon_2$ is built up from only  $\lesssim 1$ collisions per particle
in the kinetic theory. This is clearly consistent with our above classification of the range $\hat \gamma \leq 1.5$ as 
characterizing systems for which $v_2$ is built up as a small perturbation to free-streaming. 
Remarkably, Fig.~\ref{fig1} indicates that between one third and one half of the signal strength attained for an almost completely hydrodynamized large system of $\hat\gamma = 6$ can be built
up in such a much smaller non-hydrodynamized systems characterized by $\hat\gamma \leq 1-1.5$. This supports the qualitative idea that very small collisions, such as pp or pA, may build up 
a sizeable fraction of the signal strength $v_2/\epsilon_2$ seen in fully hydrodynamized large collision systems while still operating close to the free-streaming limit. 

{\bf Matching kinetic theory to viscous fluid dynamics.} 
To the extent to which collision systems hydrodynamize, one may consider describing their late-time evolution with viscous fluid dynamics from a switching time $\tau_s$ onwards. 
In the phenomenological practice of extracting $\eta/s$ from data on $v_n$, this matching of pre-hydrodynamic evolution (not necessarily given by full kinetic theory) is an important step in 
fluid dynamic models.  Its uncertainty has been quantified for large collision systems which are known to hydrodynamize~\cite{vanderSchee:2013pia,Liu:2015nwa,Chattopadhyay:2017bjs,Kurkela:2018wud,Kurkela:2018vqr}. 
As we have seen here that smaller collision systems hydrodynamize to a lesser degree, the accuracy of this matching needs to be reassessed as a function of system size, which the calculation
of the full kinetic solution allows us to do. To this end, we introduce now the viscous fluid dynamics, to which we match:
We parallel the set-up of massless transport theory by considering a conformally symmetric system with 
$\varepsilon = 3p$. The tensor decomposition $T^{\mu\nu} = \varepsilon \left( u^\mu u^\nu + \frac{1}{3} \Delta^{\mu\nu}\right) + \Pi^{\mu\nu}$ defines the local rest frame $u^\mu$, energy density $\varepsilon$ and the shear viscous tensor $\Pi^{\mu\nu}$. To set the initial values of these fluid dynamic fields at the switching time $\tau_s$, we match this tensor decomposition at $\tau_s$  to the 
energy-momentum tensor calculated from the distribution (\ref{eq4}) evolved up to $\tau_s$ with the full kinetic theory, and with $\gamma$ setting 
the kinetic viscosity in fluid dynamics. From time $\tau_s$ onwards, these fluid dynamic fields are then evolved with the Israel-Stewart viscous fluid dynamic equations 
\begin{eqnarray}
 && D\varepsilon + \left(\varepsilon + p\right) \nabla_\mu u^\mu + \Pi_{\mu\nu} \Delta^{\mu\alpha} \nabla_\alpha u^\nu = 0\, ,\label{eq5} \\
&&  \left(\varepsilon + p\right) Du^\alpha +\Delta^{\alpha\beta} \nabla_{\beta} p + {\Delta^\alpha}_\nu \nabla_\mu \Pi^{\mu\nu} = 0\, ,\label{eq6}\\
&& \tau_{\pi,{IS}}\! \left( D \Pi^{\mu\nu}\!\! +\! \textstyle{\frac{4}{3}} \Pi^{\mu\nu} \nabla_\alpha u^\alpha\!  \right)\! = - \left(\Pi^{\mu\nu} \! +\! 2 \eta \sigma^{\mu\nu} \right) . \label{eq7}
\end{eqnarray}
Here, $\Delta^{\mu\nu}=u^\mu u^\nu + g^{\mu\nu}$ is the projector on the subspace orthogonal to the flow field, $\nabla_\mu$ is 
the covariant derivative and $D \equiv u^\mu \nabla_\mu$ is the comoving time derivative. Eqs.~(\ref{eq5}) and (\ref{eq6}) result from 
energy and momentum conservation $\nabla_\mu T^{\mu\nu} = 0$, respectively. Eq.~(\ref{eq7})  ensures
for a conformal system~\cite{Baier:2007ix} that within the shear relaxation 
time $\tau_{\pi,IS}$, the shear viscous tensor relaxes to its Navier-Stokes value $-2 \eta \sigma^{\mu\nu} $, where $\eta$ is the shear viscosity and $ \sigma^{\mu\nu} = \textstyle{\frac{1}{2}} \left(\Delta^{\mu\alpha} \nabla_\alpha u^\nu  + \Delta^{\nu\alpha} \nabla_\alpha u^\mu \right)  -  \textstyle{\frac{1}{3}}   \Delta^{\mu\nu}  \nabla_\alpha u^\alpha$. {We use the second order transport coefficient $\tau_{\pi}$ to set the Israel-Stewart relaxation time $\tau_{\pi,IS}= \tau_{\pi} = 5\frac{\eta}{sT}= (\gamma \varepsilon^{1/4})^{-1}$}.

In practice, we linearize~\cite{Floerchinger:2013rya,Floerchinger:2013hza} eqs.~(\ref{eq5})-(\ref{eq7}) with respect to small eccentricity perturbations on top of an azimuthally symmetric 
background, $\varepsilon = \varepsilon_{\rm BG} + \delta \varepsilon$, $u^\mu = u^\mu_{\rm BG} + \delta u^\mu$, 
$\Pi^{\mu\nu} = \Pi^{\mu\nu}_{\rm BG} + \delta \Pi^{\mu\nu}$. After harmonic decomposition, this leads to a coupled set 
of evolution equations for 10 $\tau$- and $r$-dependent fluid field components, namely four background
field components and six components of second harmonic perturbations. This linearized treatment is sufficient to obtain exact results for the response coefficients $v_2/\epsilon_2$ studied here.
The initial conditions at the switching time $\tau_s$ are then evolved with a routine adapted from~\cite{Floerchinger:2013rya}. 
We calculate from the evolved fluid-dynamic 
fields the zeroth and second-order harmonics of the component $T^{0r}(\tau,r)$ of the energy-momentum tensor, 
and we determine $v_2$ from the ratio of the $r$-integrals of these 
components. The values for $v_2$ shown here are the $\tau\to\infty$  limit  of this procedure. Because of the conformal symmetry,
the elliptic momentum asymmetry extracted from viscous fluid dynamics can be shown to depend only on
two parameters,
\begin{eqnarray}\label{eq:scaling02}
	v_2 = v_2(\hat\gamma, \tau_s/R)\, .
\end{eqnarray}

{\bf Results from matching kinetic theory to viscous fluid dynamics.} 
We first determine the maximal value that $v_2/\epsilon_2$ can attain in a fluid-dynamic description.  This maximum is obtained for an ideal 
fluid-dynamic evolution that translates spatial gradients into momentum gradients without dissipative losses, and that is effective over
the maximal possible time, {\it i.e.} for initial conditions of (\ref{eq4}) with $\tau_s \rightarrow 0$. The resulting limiting value $v_2/\epsilon_2 = 0.51$ is shown as the blue dashed curve in Fig.~\ref{fig1}.
It is substantially larger than the full kinetic theory value at $\hat \gamma=6$. The full transport result in Fig.~\ref{fig1} approaches this ideal fluid-dynamic upper
bound slowly but steadily in the limit of very large transverse system size ($\hat\gamma \to \infty$). But even though we are dealing for $\hat\gamma = 6$ with an almost perfectly hydrodynamizing 
system, the system is still anisotropic and therefore, the signal strength $v_2/\epsilon_2$ remains substantially reduced compared to an ideal fluid-dynamic evolution initialized at $\tau_s = 0$. 

From the dash-dotted curves in Fig.~\ref{fig1}(a), one sees that viscous fluid dynamics, matched to the pre-hydrodynamic evolution at $\tau_s$, approaches the full kinetic theory calculation of $v_2/\epsilon_2$
smoothly for increasing $\tau_s$. So for fixed $\hat\gamma$, $v_2/\epsilon_2$ starts to quantitatively agree with full transport for $\tau_s \gtrsim R/\hat\gamma$, 
consistent with the observation in Fig.~\ref{fig2} that the constitutive equations are approximately fulfilled. 
For earlier initializations of fluid dynamics,
say $\tau_s < R/2\hat\gamma$, the signal $v_2/\epsilon_2$ is too strong. Indeed, for too early times,  $P_L^{hydro}$ turns negative, signaling a catastrophic failure of fluid dynamics, see
Fig.~\ref{fig2}. 

However, whether the matching of kinetic theory to viscous fluid dynamics is a quantitatively satisfactory approximation to full kinetic theory 
depends on the accuracy that one wants to achieve. The phenomenological challenge is to determine $\hat\gamma$ for fixed $v_2/\epsilon_2$, a task that becomes more
challenging for large $\hat\gamma$ where the $\hat\gamma$-dependence of $v_2/\epsilon_2$ becomes weak. For instance, for a fixed value $v_2/\epsilon_2 \approx 0.32$, a 
simulation using matching at $\tau_s = 0.15\, R$ would yield $\hat\gamma \approx 2$ while the truth of the full transport calculation is  $\hat\gamma \approx 4$. This illustrates
that even small uncertainties in $v_2/\epsilon_2$ for fixed $\hat\gamma$ can result in large uncertainties in extracting $\hat\gamma$ from a given $v_2/\epsilon_2$. 
As $\eta/s$ is inversely proportional to $\hat\gamma$, this poses a challenge for extracting $\eta/s$ from fluid dynamic simulation with accuracy significantly better than a factor of two.

While we have discussed so far only the use of full kinetic theory for the pre-hydrodynamic stage up to $\tau_s$, a much more approximate, simplified procedure is 
currently in phenomenological use. It consists of initializing fluid dynamics from free-streamed distributions at time $\tau_s$~\cite{Liu:2015nwa}.
This may be justified qualitatively on the grounds that both free-streaming and kinetic transport smoothen 
gradients in initial distributions and that any difference between free-streaming and transport will emerge only gradually 
at times $\tau \sim l_{\rm mfp} \sim R/\hat \gamma $ at which fluid dynamics starts 
to give a good description of the $T^{\mu\nu}$-evolution (see Fig.~\ref{fig2}).
Fig..~\ref{fig1}(b) shows that matching viscous fluid dynamics to free-streamed initial distributions comes with large uncertainties displayed by the wide spread of curves for different $\tau_s$.

{\bf In summary}, we have provided a full kinetic theory calculation of the opacity dependence of elliptic flow, ranging from systems that are sufficiently small to evolve close to free-streaming, up to systems
that are sufficiently large to exhibit fluid dynamic behavior already at times $\tau \ll R$. We find in very small systems a surprisingly rapid onset of signal strength $v_2/\epsilon_2$ with system size.
In particular, very small collision systems that allow for only up to one isotropizing large-angle scattering per particle excitation and that do not hydrodynamize significantly on time scales 
$\tau < 2 R$ are still found to build up one third to one half of the signal strength observed in almost completely hydrodynamized, large collision systems. 
That $v_2/\epsilon_2$ rises with system size most rapidly in the range up to $R < l_{\rm mfp}$ where collective flow results from perturbative (in $\hat \gamma$) corrections to free-streaming
is a characteristic feature of kinetic transport theory established here. These findings are qualitatively in line with the potentially contradictory requirements that $v_2$ attains sizeable values already
in the smallest $pp$ collision systems despite many other observations in pp collisions being seemingly consistent with an approximate free-streaming picture. 

Our study demonstrates that matching a kinetic theory pre-hydrodynamic stage at $\tau_s$ to a viscous fluid dynamic description can yield accurate results for $v_2/\epsilon_2$
if the switching time $\tau_s$ is sufficiently late and if the system is sufficiently opaque ($\hat \gamma \gg 1$). The accuracy of this matching degrades only gradually with decreasing 
$\hat \gamma$. We note that for systems for which the pre-equilibration dynamics is sufficiently short, 
 full kinetic transport may be replaced by linear response~\cite{Keegan:2016cpi,Kurkela:2018wud,Kurkela:2018vqr}, for which a numerical code K{\o}MP{\o}ST \cite{Kurkela:2018wud,Kurkela:2018vqr} is available.
However, in smaller systems, for which the pre-equilibirum dynamics needs to be followed to later times $\tau_s \sim R$, full kinetic transport is needed, and for even smaller systems $\hat\gamma < 1$ 
the single hit approximation to kinetic theory is sufficient. 

In short, we have demonstrated that while pp collisions are very different from AA collisions since they realize a close to free-streaming picture that differs qualitatively from hydrodynamics, 
the collectivity in both systems can still arise from the same microscopic interactions.

\bibliographystyle{elsarticle-num}

\end{document}